\pdfoutput=1
\documentclass[nonacm]{acmart}
\usepackage{comment}
\usepackage{multirow}
\usepackage{tikz}
\usepackage{ctable}
\usepackage{kotex}

\newcommand{\sys}{ModSandbox}

\newcommand{\BC}[1]{{%
\begin{tikzpicture}[baseline=(C.base)]
\node[draw,circle,fill=black,inner sep=0.2pt](C) {\textcolor{white}{#1}};
\end{tikzpicture}
}}

\AtBeginDocument{%
  \providecommand\BibTeX{{%
    \normalfont B\kern-0.5em{\scshape i\kern-0.25em b}\kern-0.8em\TeX}}}

\begin{document}

\title[Online Community Moderation Through Error Prediction and Improvement of Automated Rules]{\sys{}: Facilitating Online Community Moderation Through Error Prediction and Improvement of Automated Rules
}

\author{Jean Y. Song}
\authornote{Both authors contributed equally to this research.}
\email{jeansong@dgist.ac.kr}
\orcid{0000-0003-4379-3971}
\affiliation{%
  \institution{DGIST}
  \city{Daegu}
  \country{Republic of Korea}
}
\author{Sangwook Lee}
\authornotemark[1]
\email{sangwooklee@kaist.ac.kr}
\affiliation{%
  \institution{KAIST}
  \city{Daejeon}
  \country{Rupublic of Korea}
}

\author{Jisoo Lee}
\affiliation{%
  \institution{Beeble Inc.}
  \city{Seoul}
  \country{Repulic of Korea}
}
\email{jisoo.lee@beeble.ai}

\author{Mina Kim}
\affiliation{%
  \institution{Kakao Corp.}
  \city{Jeju}
  \country{Republic of Korea}
}
\email{iamhappy537@gmail.com}

\author{Juho Kim}
\affiliation{%
  \institution{KAIST}
  \city{Daejeon}
  \country{Republic of Korea}
}
\email{juhokim@kaist.ac.kr}

\renewcommand{\shortauthors}{Song and Lee, et al.}

\begin{abstract} %
Despite the common use of rule-based tools for online content moderation, human moderators still spend a lot of time monitoring them to ensure that they work as intended. Based on surveys and interviews with Reddit moderators who use AutoModerator, we identified the main challenges in reducing false positives and false negatives of automated rules: not being able to estimate the actual effect of a rule in advance and having difficulty figuring out how the rules should be updated. To address these issues, we built \sys{}, a novel virtual sandbox system that detects possible false positives and false negatives of a rule to be improved and visualizes which part of the rule is causing issues. We conducted a user study with online content moderators, finding that \sys{} can support quickly finding possible false positives and false negatives of automated rules and guide moderators to update those to reduce future errors. %

\end{abstract}



\begin{teaserfigure}
  \centering
  \includegraphics[width=1.0\textwidth]{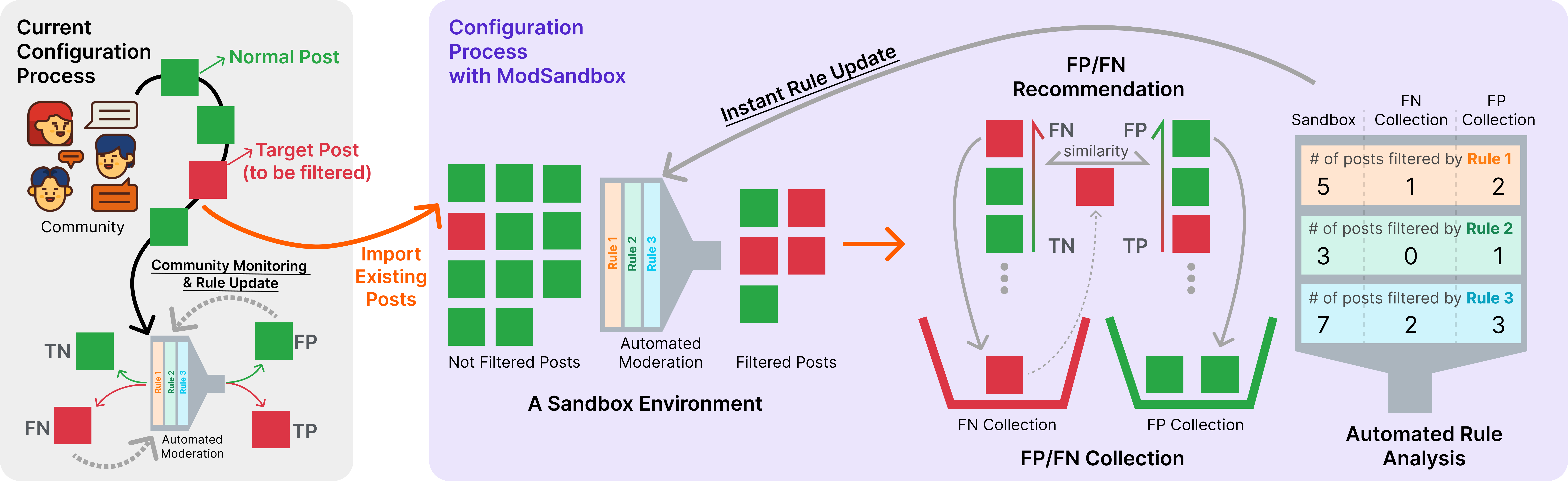}
  \caption{\sys{} supports online community moderators with error prediction and improvement of their automated rules. The moderators currently monitor their community and update the rules to catch the target posts (posts they want to filter) based on their previous experience with false positives and negatives. \sys{} provides features to help predict possible false positives and false negatives using existing posts (\textbf{A Sandbox Environment} and \textbf{FP/FN Recommendation}), and to improve automated rules (\textbf{FP/FN Collection} and \textbf{Automated Rule Analysis}).}
  \Description{The teaser image of our paper. This image describes the current automated rule configuration process and one with ModSandbox}
  \label{fig:teaser}
\end{teaserfigure}

\maketitle
\section{Introduction} \label{sec:section1_new}

Communities on social platforms such as Reddit, Discord, and Twitch have a group of users who volunteer to moderate their communities, called online moderators~\cite{matias2016going, jiang2019moderation}. They respond to the behavior of community members that violate rules and work to improve overall interaction experiences between community members~\cite{grimmelmann2015virtues, seering2019moderator}. %
Contrary to large social platform companies like Facebook and Twitter that apply machine learning algorithms to regulate user generated content at scale~\cite{gorwa2020algorithmic}, small online community moderators prefer using rule-based tools like a keyword filter because they can customize the programmed conditions to suit their community's needs, and the tool's behaviors are more predictable and interpretable. This helps the moderators apply community-specific norms and transparently explain the tool's malfunction to the members when it happens~\cite{kiene2016surviving,jhaver2019human}. 

These rule-based automated tools monitor posts being uploaded in real time and remove, hide, tag, and comment on posts based on their programmed conditions~\cite{jhaver2019human}. 
For example, Reddit moderators use AutoModerator, a site-wide rule-based moderation tool that can automate tedious moderation tasks~\cite{kiene2016surviving, jhaver2019human}. According to the Reddit Transparency Report 2021, AutoModerator removed 58.9\% of the removed Reddit content~\cite{reddit2021transparency}. Discord moderators use various third-party moderation bots with rule-based moderation functions, such as word filters and user ban lists~\cite{kiene2020uses, kiene2019technological}, which more than 18.1M Discord servers use~\cite{discordbot}.

However, oftentimes a moderation tool may not work as intended -- missing posts that the moderators wanted to catch (false negative) or catching the posts that the moderators did not want to catch (false positive). These errors adversely affect the community and require additional work from the moderators. For example, if the automated tool does not immediately remove hateful speech, it can increase the level of emotional stress in the community~\cite{saha2019prevalence}. On the other hand, if it removes an innocent post, it can cause a backlash from the authors due to being seen as censorship~\cite{jhaver2019did,diakopoulos2016accountability}. This may decrease user engagement and cause them to leave the community~\cite{juneja2020through,Collier_2014,AlfonsoIII_2014}. %
To resolve false negatives and false positives, moderators remove or approve posts manually~\cite{seering2019moderator}, or update their automated rules as an afterthought to prevent future problems~\cite{jhaver2019human}. %
We believe that a better solution would be to predict possible false negatives and false positives \emph{beforehand} so that moderators can minimize the errors of automated rules before their deployment.

To understand the challenges with configuring automated rules, 
we conducted surveys and a round of interviews with volunteer moderators on Reddit who actively use AutoModerator. %
From in-depth interviews with five Reddit moderators, we found four main challenges moderators encounter during a typical moderation process:  %
1) there is no way to estimate the actual effects of a rule in advance, 2) it is hard to detect false positives of a rule after its deployment, 3) it is hard to figure out how the rule should be updated to reduce false positives and false negatives, and 4) it is hard to understand which part of the rule is causing a problem. 

Based on the identified challenges, 
we built \sys{}, a sandbox system where %
moderators can test their automated rules by using existing community posts before the actual deployment of rules.
\sys{} has four main features corresponding to the identified challenges: 1) a sandbox to enable prompt configuration evaluation without affecting the actual posts and comments, 2) a recommendation of possible false positive and false negative posts to enable faster discovery, , 3) a temporary repository feature to allow users to collect actual false positive or false negative posts to identify the common patterns in them, and 4) a visualization to analyze how the rule affects the posts. \sys{} uses a machine learning-based sentence encoder to calculate the possibility of false positive and false negative for each post imported into the system.

We conducted a user study with 10 active online moderators to assess whether and how \sys{} helps the configuration process of an automated moderation tool. \sys{} was able to sort posts in a sandbox for the participants to easily detect false positives and false negatives. Also, with \sys{}, participants wrote more sophisticated rules that can filter target posts more precisely. We observed that the participant tried to improve their automated rules with structured and iterative processes using \sys{} features. Finally, we compared their perceived usefulness scores of \sys{} and its features according to the types of task and the user's condition to highlight their strengths and weaknesses.

We conclude our work by discussing how the proposed design of a system can be improved, potentially facilitate distributed governance for online communities, and how it can reduce cognitive labor in setting up automated moderation tools.

\section{Related Work} \label{sec:section2_new}

We focus our review on automated content moderation on social platforms and designing systems for online content moderation. In addition, we provide background information on Reddit's AutoModerator, which we use for our user study evaluation. 

\subsection{Automated Content Moderation on Social Platforms}

There are two levels of content moderation on social platforms: community-level moderation led by users and platform-level moderation led by platform companies~\cite{seering2020reconsidering}. Social platforms such as Meta and Twitter employ paid workers to find and remove content that violates site policies %
~\cite{roberts2019behind}. They focus on policing harmful behaviors such as spreading fake news and hate speech~\cite{kumar2021battling, hatefultwitter}, sharing unhealthy tags~\cite{chancellor2016thyghgapp}, posting violent or sexual content and using slurs and swear words. Recently, many platforms have adopted machine learning-based systems to automatically manage their content at scale~\cite{gorwa2020algorithmic}. For example, Facebook uses algorithms to automatically suspend accounts that do not use real names. However, Facebook had to update its algorithm regarding the real name policy because its definition of real name did not include Native Americans who have last names such as ``Lone Hill'' or ``Brown Eyes''~\cite{vaccaro2020end}. We note that one down side of machine learning-based content moderation is that it lacks context, having the possibility to exclude or disadvantage minor groups or small communities.

Other social platforms such as Reddit and Discord allow voluntary moderators to manage their communities themselves%
~\cite{matias2016going}. 
Typically, these moderators are elected among community members who understand the community norm or are invited by other moderators~\cite{seering2019moderator}. Unlike paid workers on large platforms who do not have the authority to decide or change the policy of the platform, voluntary moderators are deeply involved in establishing, determining, and executing their community rules~\cite{seering2019moderator}. %
Although the voluntary moderation opportunity increases the degree of freedom that moderators have in applying the rules for their particular community, it requires moderators to spend a lot of their time and effort on the moderation tasks. 
As voluntary moderators cannot spend most of their time monitoring their communities, many adopt moderation tools provided by the platform~\cite{jhaver2019human}, third-party companies~\cite{cai2019categorizing}, and platform users~\cite{kiene2019technological}. These tools are mostly rule-based,  %
which allows moderators to directly control how they operate and, if necessary, to transparently communicate with community members on the cause of moderation errors, i.e. false positives and false negatives caused by automated rules~\cite{jhaver2019human}. 

Even if these rule-based moderation tools are more straightforward and flexible than machine learning-based tools, they often do not work as the moderators intended, which requires human moderators to constantly update the configurations to reflect their intention~\cite{chandrasekharan2019crossmod}. %
For example, users can use abbreviations~\cite{sood2012profanity}, intentional misspellings, and lexical variation~\cite{chancellor2016thyghgapp} of a banned word to avoid automatically being filtered. The moderators then have to update their filter by adding these variations~\cite{jhaver2019human}. While these false negatives are dealt with by updating the rules, false positives are more annoying because they require moderators to manually reverse each issue~\cite{seering2019moderator}.
In this study, we explore the effectiveness of a moderation support system that allows its users to predict false positives and false negatives of a rule-based automated content moderation tool that is applied to the content of their own community so that moderators can improve their rules to prevent future false positives and false negatives.

\subsection{Designing a System for Content Moderation}

In the context of online content moderation, many studies have introduced machine learning-based classifiers to detect harmful comments and malicious users in the online space. Types of classifier include the detection of cyberbullying~\cite{dinakar2011modeling}, profanities and insults~\cite{sood2012automatic}, pornographic content~\cite{singh2016behavioral}, hate speech~\cite{davidson2017automated}, and abusive behaviors~\cite{nobata2016abusive, chandrasekharan2017bag}. As machine learning techniques evolve, recent studies made classifiers multimodal and community-specific, so that the classifiers can reflect each community's preference and culture. For example, Chancellor et al.~\cite{chancellor2017multimodal} developed a multimodal classification model to detect images and text that promote eating disorders, which do not fall into the traditional category of harmful content. Furthermore, Chandrasekharan et al.~\cite{chandrasekharan2019crossmod} trained macro norms and community-specific %
to make the classifier more suitable for each community. 
While previous studies have focused on supervised learning to classify behaviors generally considered harmful, our study adopts an embedding model pretrained by a large language corpus to find comments with few examples that represent the individual moderator's intention. Our system combines the filtering results and their semantic similarities with examples to find possible false positives and false negatives.

Although there have been studies on algorithmic support for online content moderation, few studies have proposed to visualize actual content of the community, such as comments to help configure automated rules and support the moderation process.
CommentIQ~\cite{park2016supporting} is an interactive visualization tool for online news comment moderators, which helps to find high-quality comments for readers. The user can filter the comments by criteria, location, and times by brushing and linking on their distribution visualization. Also, the system allows the users to reflect their preference to high-quality comments into the sorting order by setting the weights for predefined criteria. 
Recently, FilterBuddy~\cite{jhaver2022designing} introduced a tool for YouTube creators to help moderate comments on their videos. The user could customize the word filters to hide or remove comments with specific words in existing filter lists. The system used existing comments to show what and how many comments were filtered, to help evaluate the performance of the filter. In this work, we focus on system design to help community moderators configure a rule-based automated tool that supports combinations of word filters to find posts that violate community rules. Our system shows the expected results of the configured tool using existing posts in a real community and visualizes the relationship between the posts and the configuration to help users analyze each filter.

\subsection{Background: Reddit AutoModerator}

Reddit AutoModerator is a rule-based automated moderation tool developed by one of the Reddit moderators, Chad Birch, in 2013~\cite{jhaver2019human}. By configuring AutoModerator using YAML, Reddit moderators can create their own automated rules suitable for each subreddit's preference and culture. In 2015, Reddit officially integrated AutoModerator into the platform as a feature of the default moderation tools. According to Reddit transparency reports\cite{reddit2021transparency}, AutoModerator removed about 103.6M content in 2021, which is 20.9\% more than 2020, and 58.9\% of all content removed by moderators.

AutoModerator works on all the posts and comments on a subreddit according to the automated rules, which a human moderator last saved in their AutoModerator. In other words, once moderators change their rules, AutoModerator applies the change to newly uploaded contents, not the previous ones. Most moderators write multiple automated rules to detect profanity, slurs, and a set of posts that violate specific rules of an individual subreddit. Each rule has one or more checks and actions. The check consists of a field that AutoModerator reviews and a list of keywords, phrases, and regular expressions. The tool verifies whether the fields, such as title and body, include any words and phrases or match with regular expressions in the list. 
The check also supports verifying content length, the number of user reports, the account age, reputation score, and other features of the Reddit post. A human moderator can combine multiple checks to fine-tune the scope of the rule. The rule also includes actions that indicate the moderation action to perform against the posts identified by the check. %

\section{Interview: Challenges Encountered During Configuration Process}
\label{sec:section3_new}

\begin{table}[]
\resizebox{\textwidth}{!}{
\begin{tabular}{cccc|cc}
\textbf{No.} &
  \textbf{Age} &
  \textbf{Moderator Periods} &
  \textbf{Gender} &
  \textbf{AutoModerator Knowledge} &
  \textbf{Configure AutoModerator?} \\ \hline
P1 &
  35-44 &
  6 months &
  M &
  \begin{tabular}[c]{@{}c@{}}Yes, I'm not an expert \\ but I know enough to use in my own sub\end{tabular} &
  Yes, occasionally \\ \hline
P2 & 35-44 & 4 years & M & Yes, I'm an expert                & Yes, most of the time \\ \hline
P3 &
  35-44 &
  3 years &
  M &
  \begin{tabular}[c]{@{}c@{}}Yes, I'm not an expert \\ but I know enough to use in my own sub\end{tabular} &
  Yes, most of the time \\ \hline
P4 & 18-24 & 2 years & M & Yes, I'm an expert                & Yes, most of the time \\ \hline
P5 & 18-24 & 5 years & M & Well, I think I know a little bit & Yes, most of the time \\ \hline

\end{tabular}}
\caption{Background Information of Interview Participants}
  \label{tbl:interview_participants}
  \Description{The table shows the background information (age, moderator periods, gender, AutoModerator knowledge, and AutoModerator configuration experience) of interview participants }
\end{table}

To reflect the current practices and challenges of configuring AutoModerator into the design of our system, we conducted semi-structured interviews with five Reddit moderators (Table~\ref{tbl:interview_participants}) who have experience configuring AutoModerator. 
To recruit AutoModerator users among Reddit moderators, we sent online survey links to moderators selected from a list of popular subreddits~\footnote{https://www.reddit.com/r/ListOfSubreddits/wiki/listofsubreddits/} through the internal Reddit mailing system. A total of 50 moderators answered the online survey that asked for knowledge of how to configure AutoModerator and whether they configured AutoModerator themselves. Then, we sent interview recruitment emails to survey respondents who responded that they have configured AutoModerator by themselves occasionally or most of the time, and left their email addresses for a further in-depth interview.

Each interview session lasted 40-70 minutes through an online conference call and each participant was paid a \$30 Amazon gift card for their participation. To extract the challenges of the configuration process from the interview transcriptions, four authors and one assistant participated in an iterative coding process through multiple pairing sessions. The authors were randomly paired for each session and coded an interview transcription. We immediately resolved any disagreement through discussion. After coding all five transcriptions, the authors gathered for four consecutive two-hour meetings to interpret and find patterns in the code and discussed until a consensus on the final codebook was reached on derived themes from the process. %

According to interviews, their configuration process to update an automated moderation tool could be divided into two steps: error identification step and rule update step to avoid similar errors in the future. In the following, we describe the four challenges (C1-C4) that online community moderators face in each step.

\begin{figure}[t]
    \centering
    \includegraphics[width=\textwidth]{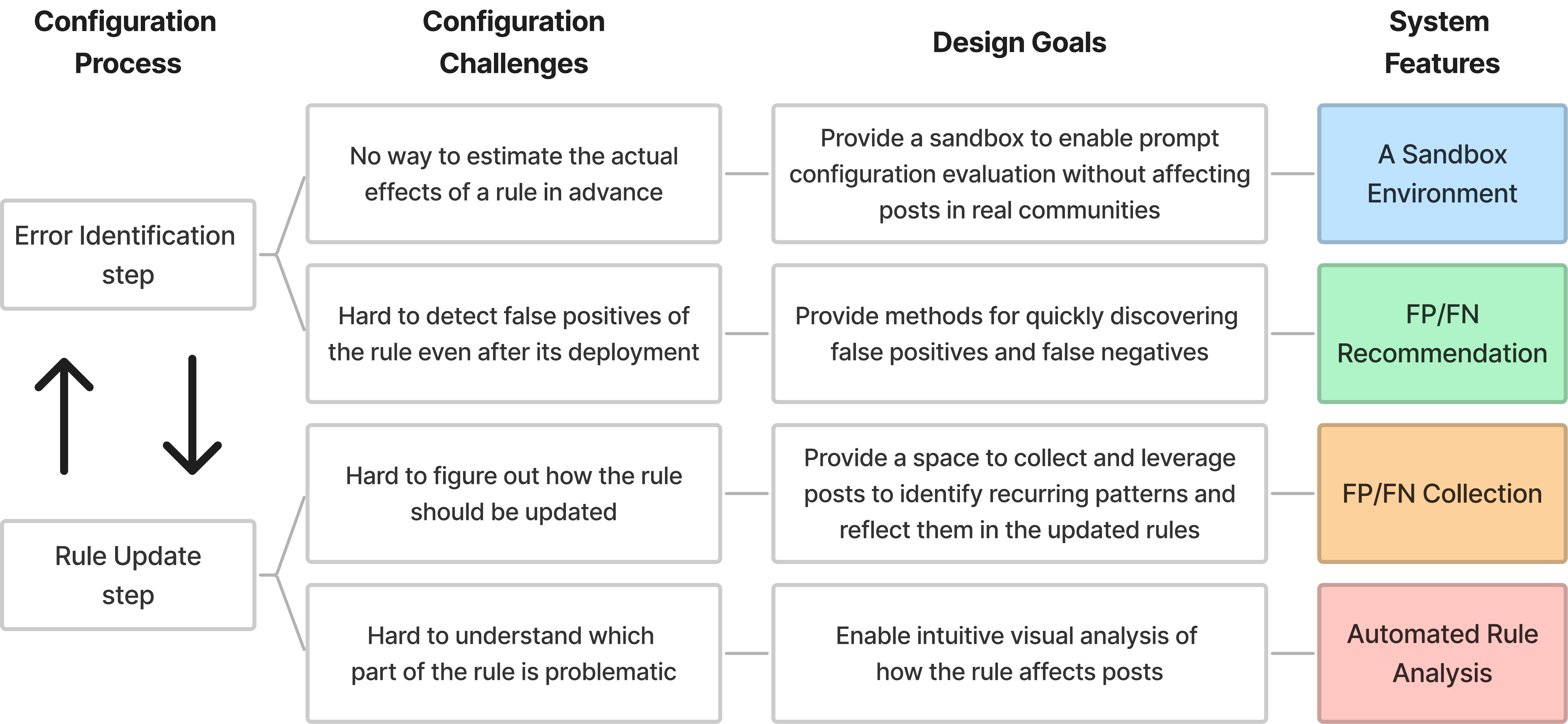}
    \caption{
        A diagram that shows relationship between configuration process, challenges, design goals, and system features.
    }
    \Description{This figure shows relationship between configuration process, challenges, design goals, and system features.}
    \label{fig:system_design}
\end{figure}

\paragraph{C1. No way to estimate the actual effects of a rule in advance}
When moderators want to discern errors from AutoModerator they configured, they cannot estimate the actual effects of their rules in advance. Participants said that they monitor their community and mod tools such as moderation queue, moderation logs, and Modmail to check for errors that have already occurred in their communities. The moderation queue shows the posts or comments reported by users, letting moderators notice the posts AutoModerator missed. The moderation logs and Modmail (internal mailing system for Reddit Moderators) help moderators find false positives by showing the operation history of AutoModerator and user's claim, respectively. However, none of them supported checking automated rules in advance before AutoModerator affects community posts. P1 complained that there is no testing protocol to ensure that it works in the real world. 
P1 and P4 reported that they use fake accounts to submit test posts in their community to check the operation of AutoModerator. However, they can test AutoModerator with only a few imaginary posts that poorly represent real-world posts. Thus, moderators face difficulty in estimating examples of possible false positives and the actual effect of AutoModerator on the community. 

\paragraph{C2. Hard to detect false positives of the rule even after its deployment}
Although moderators can search for false positives in Modmail or moderation log, they have difficulty finding false positives through those mod tools. If a user's post is removed without violating any rules, the user can appeal to moderators through Modmail. Then, the moderators can review their removed posts, which allows them to discover an issue with the AutoModerator configuration. Alternatively, moderators can detect an issue while regularly reviewing the moderation logs, where all history of moderators' actions including AutoModerator's is saved. However, Modmail requires users to claim innocent removal of their posts, which inevitably leads to many latent false positives. Furthermore, Reddit does not have an official and individual appeal process for users whose posts are removed by AutoModerator~\cite{juneja2020through}, letting users give up appealing the removal due to the inconvenience of the process. On the other hand, checking the moderation log is inefficient and can be mentally harmful to moderators. %
P1 said ``The harder part is always posts that do get moderated as opposed to posts that don't get moderated. [...] I won't, you know, necessarily see them by default unless I go searching in the automoderator log. Then, which I don't really do that often.''

\paragraph{C3. Hard to figure out how the rule should be updated} \label{sec:interview_rule}
When moderators update automated rules to prevent identified errors, they tend to narrow down the rule by finding additional patterns of target posts for a new rule, check, and strings, which can be difficult for novice moderators. P3 said that they tried to narrow down a rule as much as possible to keep it effective because they did not want to remove anything innocent. P2 said that updating the rules is just pattern recognition to spot the difference between the ones that were good and bad. 
He tended to make more complex rules to be more precise. However, this way of thinking requires recalling the errors that moderators identified during the configuration process and finding patterns that can be represented in the form of AutoModerator rules. Therefore, it can be difficult for novice moderators who lack experience with manual moderation and AutoModerator.

\paragraph{C4. Hard to understand which part of the rule is problematic}
Moderators reported difficulty in debugging their rules. In our interview, the participants shared how they update automated rules to avoid the recurrence of the same error. Since AutoModerator can work with multiple rules, they first identify a rule that generates the error among the AutoModerator configuration. When they find the rule that catches innocent posts, they tend to eliminate a check or a keyword to avoid further false positives. However, two participants (P1, P4) responded that it is difficult to understand which part of the content triggers which rule and vice versa.
P4 said, ``There's a few times that it's picked out comments that I can't figure out what's the word. Every so often.'' In addition, he shared how he uses the action reason feature of AutoModerator. Moderators can write a rule with different action reasons that are displayed with actions in the moderation log to notice which rule was involved in the action. However, this feature does not support highlighting which part of the rule was involved and requires additional labor to manage the reasons in the automated rules. 
\section {\sys{}: System Design} \label{sec:section4_new}

This section describes our system's design goals, which we set to resolve the challenges that are identified from the interviews. We then introduce \sys{} (Figure~\ref{fig:system}), a sandbox environment that is built to support online community moderators to easily predict false positives and false negatives and update their automated rules to reduce them.

\subsection{Design Goals}

We set two high-level goals in designing our \sys{} system as follows:

\begin{itemize}
\item Help moderators quickly find possible false positives and false negatives (in accordance with the error identification step in Figure~\ref{fig:system_design}).
\item Help moderators configure more sophisticated automated rules to reduce false positives and false negatives (in accordance with the automated rule update step in Figure~\ref{fig:system_design}).
\end{itemize}

For each high-level goal, we present two specific design goals and how they can resolve the four challenges found in Section~\ref{sec:section3_new} (see also Figure~\ref{fig:system_design}).

\subsubsection{DG1. Provide a sandbox to enable prompt configuration evaluation without affecting posts in real communities.}

According to our surveys and interview study, moderators do not have a way to estimate the results of an updated automated tool in the real world. A sandbox environment can be a solution, which imports real posts from the moderator's community and helps moderators evaluate the automated rules in a simulated environment without affecting posts and comments in their real community. %

\subsubsection{DG2. Provide methods for quickly discovering false positives and false negatives.}

Interviewees reported that it is hard to recognize false positive posts unless they are reported by users because often they are buried within other posts in moderation logs. %
To address this, natural language processing (NLP) techniques can be used to help moderators quickly spot false positives and false negatives and resolve errors in their rules. %

\subsubsection{DG3. Provide a space to collect and leverage posts to identify recurring patterns.}

Providing a space for moderators to collect and leverage posts such as false positives and false negatives would reduce the cognitive load in finding patterns from them. When moderators try to update automated rules, they need to find patterns of recurring errors and reflect them into updated rules. Specifically, they tend to find common features of false positives and negatives they observed during moderation to expand or narrow down the condition of their rules. Thus, we proposed the feature of collecting false positives and negatives to discover their patterns. %

\subsubsection{DG4. Enable intuitive visual analysis of how the rule affects posts.}

We suggest providing visual support to help analyze which part of a post caused the automated rule to filter it and how many posts are affected. We found that moderators struggle to recognize the relationship between automated rules and affected posts. %
In a previous study, Jhaver et al.~\cite{jhaver2019human} also discussed that visualization of the effect of each rule can help moderators configure their automated rules.

\begin{figure}[t]
    \centering
    \includegraphics[width=\textwidth]{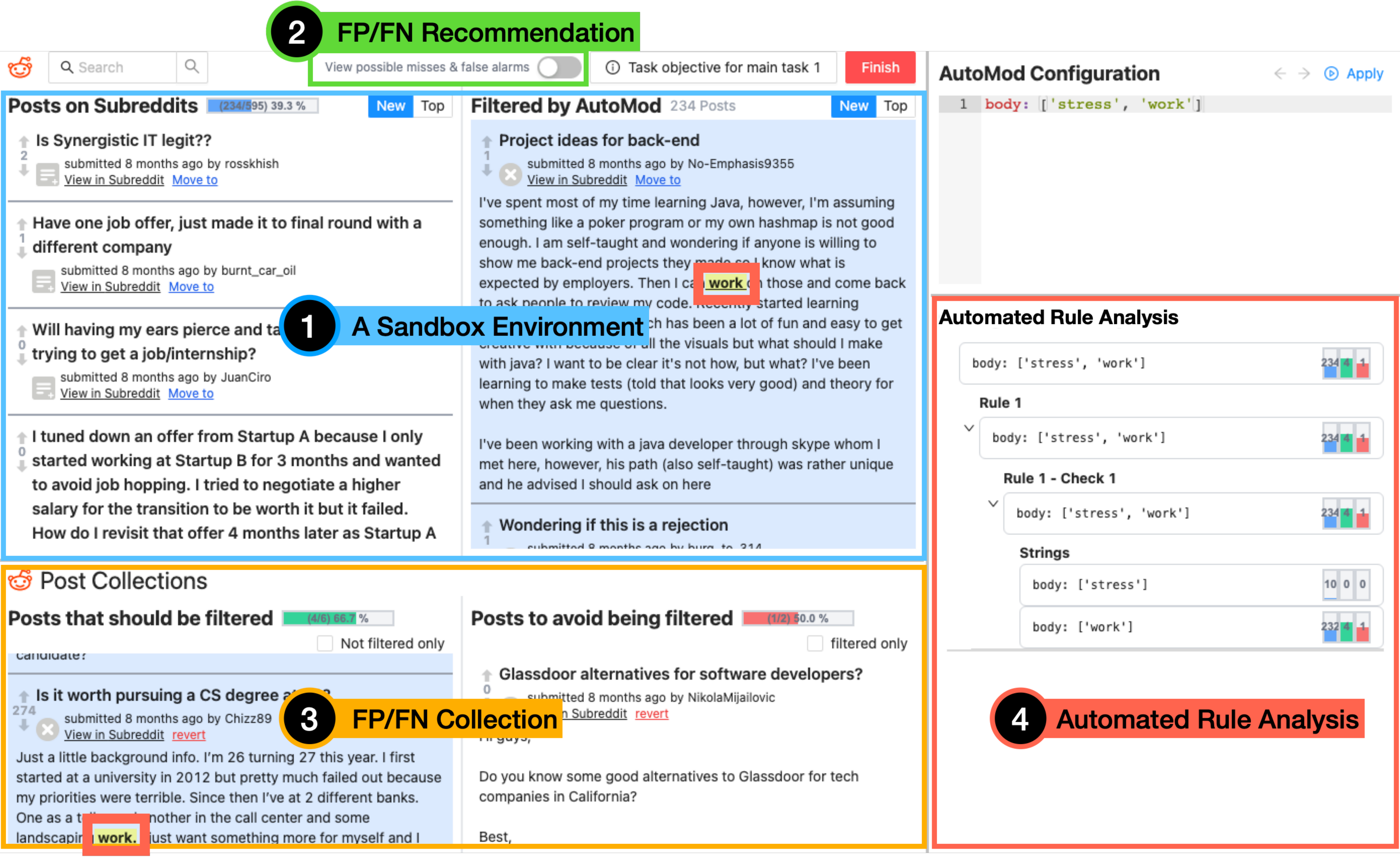}
    \caption{
        An overview of the four main features of \sys{}.  \protect\BC{1} is a ``Sandbox Environment'' where a moderator can import all the posts from their community. \protect\BC{2} is a toggle button that rearranges the posts in the sandbox area from the most ``Possible misses and false alarms'' to the least. It helps moderators to more quickly find possible misses (false negative) and false alarms (false positives). \protect\BC{3} is the ``FP/FN Collection'' area that helps moderators to collect interesting posts for finding their patterns for further rule updates. \protect\BC{4} is the ``Configuration Analysis'' panel that helps analyze how the rule affected the posts in the sandbox. It shows the number of filtered posts in ``Sandbox Environment'' and ``Post Collections'' (FP/FN Collection) with color bars and highlight the part of those filtered posts in their panels (red boxes in \protect\BC{1}, \protect\BC{2}) for macro and micro-level support of debugging each configuration.
    }
    \Description{This figure shows the overview of \sys{} interface and highlights part of the interface to indicate where each feature is.}
    \label{fig:system}
\end{figure}

\begin{figure}[t]
    \centering
    \includegraphics[width=\textwidth]{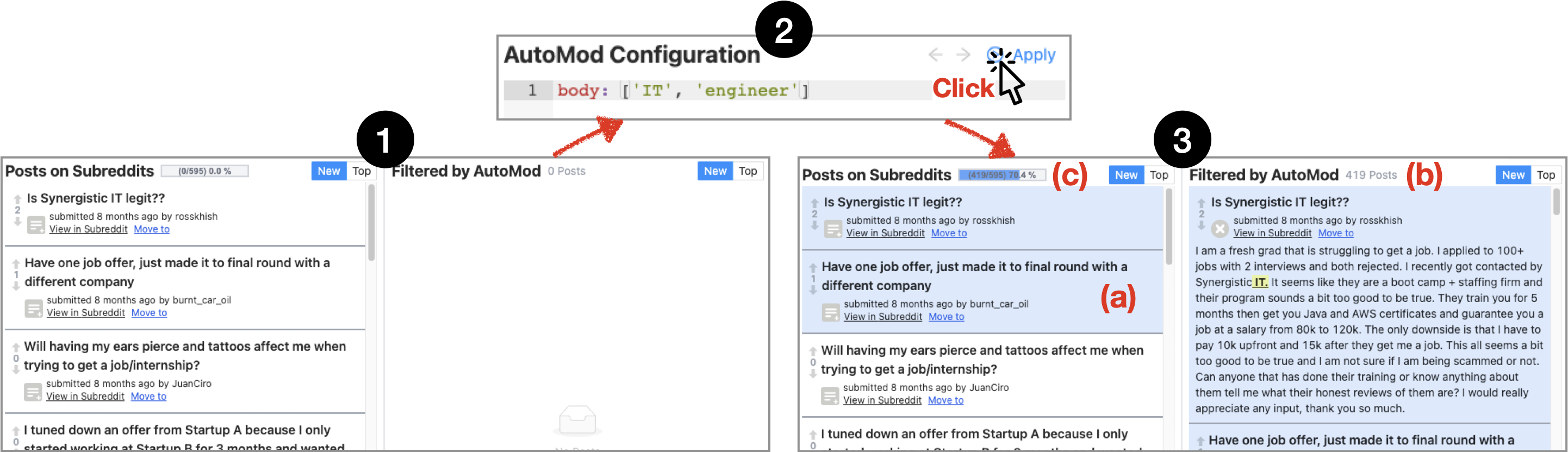}
    \caption{
        Shows how to use a Sandbox Environment. \protect\BC{1} shows the sandbox right after importing posts from a community. When a user clicks on the ``Apply'' button after writing their rules in the \protect\BC{2} ``AutoMod Configuration'' panel, \protect\BC{3}-(a) the background turns blue for the posts filtered by the rules, \protect\BC{3}-(b) ``Filtered by AutoMod'' gathers them in a separate panel for easy browsing, and \protect\BC{3}-(c) a blue bar graph shows the ratio of filtered posts to imported posts.%
    }
    \Description{This figure shows the process how the automated rules apply to posts imported in a sandbox environment}
    \label{fig:sandbox}
\end{figure}

\subsection {Feature 1: A Sandbox Environment}

\sys{} provides moderators with an isolated sandbox environment (\BC{1} in Figure~\ref{fig:system}), which allows moderators to virtually test their automated rules on posts that already exist in their communities and their moderation logs. The sandbox helps moderators identify or predict any issues with the rules without affecting the posts in their actual communities.  
Figure~\ref{fig:sandbox} shows an example of the use of our sandbox environment. As an example, a moderator imports posts from a subreddit named r/cscareerquestions into the panel ``Posts on Subreddits''. Then they write an automated rule in the ``AutoMod Configuration'' panel to filter any post with the words `IT' and `engineer'. After they click on the ``Apply'' button, every post that includes the keywords turns blue to provide a visual comparison between filtered and not-filtered posts. 
Moderators can also see the filtered posts in the ``Filtered by AutoMod'' panel, which gathers them in one place for easy browsing. 
This rearrangement and coloring of posts help moderators understand which types of posts are affected by the rule. Additionally, a horizontal bar right next to the word ``Posts on Subreddits'' presents the ratio of filtered posts. In the figure, we can observe that more than 70\% of the posts include the two keywords. Removing posts with these keywords may be a bad idea because it removes most of the posts in the community. That is, this ratio bar helps moderators understand the effect of automated rules so that they can assess whether the rules work as intended or harm the community. 

\subsection {Feature 2: FP/FN Recommendation}

\begin{figure}[t]
    \centering
    \includegraphics[width=\textwidth]{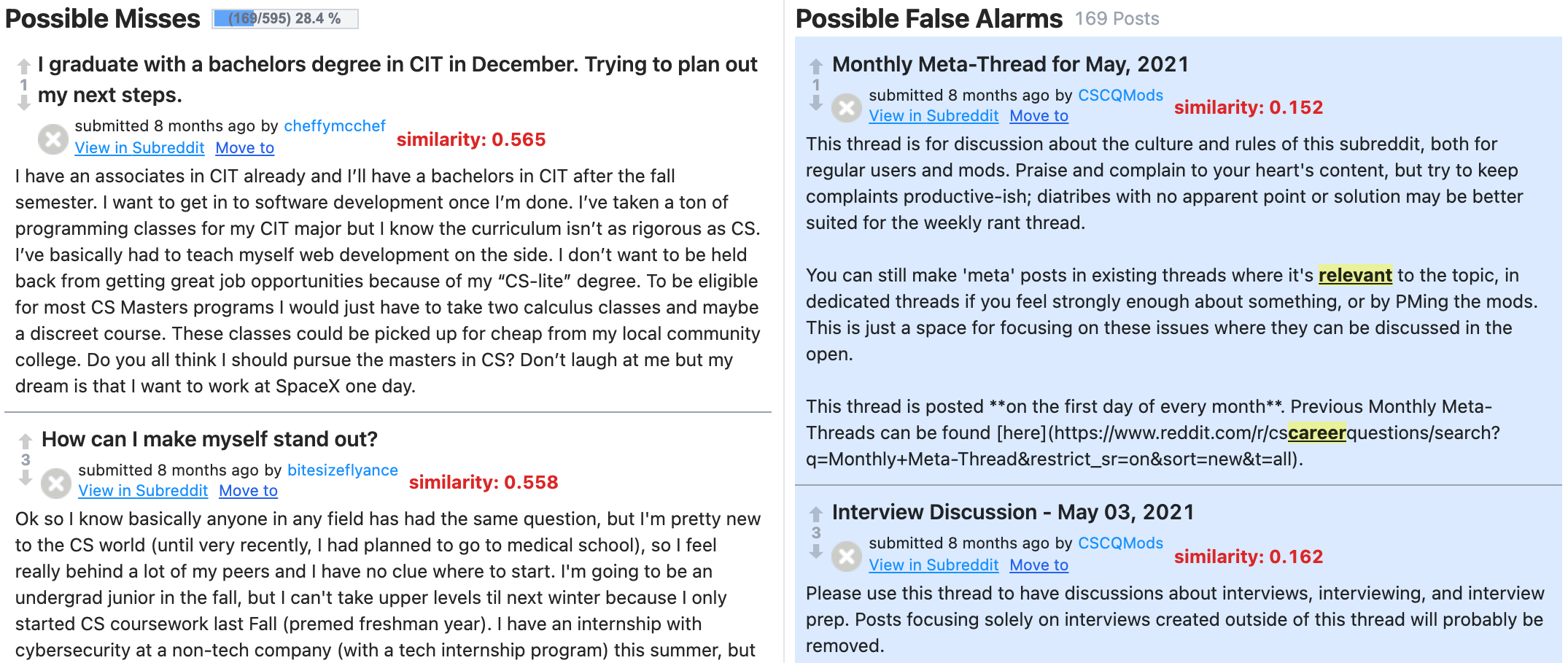}
    \caption{
        Example of possible misses (false negatives) and false alarms (false positives) of the configured rules in Task 2 of our main user study. Participants were guided to detect posts about asking whether or how to get CS-relevant jobs without CS-relevant degrees. The more probable posts that are being missed are listed at the top (e.g., similarity 0.565 is larger than 0.558), and the opposite happens for the false alarms (similarity 0.152 is smaller than 0.162). The similarity values are hidden in the actual interface.
    }
    \Description{This figure shows an example result from FP/FN recommendation with the semantic similarity scores for each post.}
    \label{fig:fpfn}
\end{figure}

\sys{} provides the ``FP/FN Recommendation'' feature to help moderators quickly find issues with their automated rules, i.e. false positives and false negatives. When moderators activate this feature by toggling a button (\BC{2} View possible misses \& false alarms in Figure~\ref{fig:system}), possible false positives (equal to False Alarms) and false negatives (equal to Misses) are presented in the order of the most probable to the least probable. This feature helps moderators quickly find possible false positives or false negatives without having to browse all the posts in the sandbox.

If a filtered post is semantically far from the posts that the moderators want to filter, it is likely to be a false positive. 
Motivated by this intuition, we treat posts that are filtered but are different from posts in ``Posts that should be filtered'' as \textit{possible false positives}. This panel is part of the ``FP/FN Collection'' feature described in Section~\ref{sec:FPFN_Collection}. On the contrary, we treated posts that are not filtered, but are similar to posts in ``Posts that should be filtered'', as \textit{possible false negatives}. 
For example, as shown in Figure~\ref{fig:fpfn}, a non-filtered post with the closest distance from the posts in ``Posts that should be filtered'' (the reference point) comes at the top of the ``Possible Misses'' panel. The farthest non-filtered post from the reference point comes at the bottom of this panel. As a result, this feature lets moderators see more critical posts first and helps them quickly find clues to update their automated rules to filter these missed posts. 

For the calculation of post similarity, we adopt a sentence embedding model to encode semantic features of each post into embedding vectors. The pretrained sentence-level embedding outperforms word-level embedding in various transfer tasks~\cite{conneau2017supervised}. Also, the sentence embedding can complement the limitation of AutoModerator's word filtering by considering the context of the post to find false positives and negatives. 
When moderators import their community posts, our system computes and saves an embedding vector for each post using Universal Sentence Encoder~\cite{cer2018universal}, one of the popular open-source sentence embedding models. Then, it computes the cosine similarities between the saved vectors and the average vector of the posts in the ``posts that should be filtered'' and sorts the possible misses and false alarms in the order of similarity. Although the vector encoding step requires a high computation cost proportional to the number of posts, this is a one-time computation that only occurs after moderator imports posts from their community. The time to calculate the cosine similarities is also proportional to the number of posts in the system, but the calculation is much faster because it does not require deep models. %

\subsection {Feature 3: FP/FN Collection}
\label{sec:FPFN_Collection}

\begin{figure}[t]
    \centering
    \includegraphics[width=\textwidth]{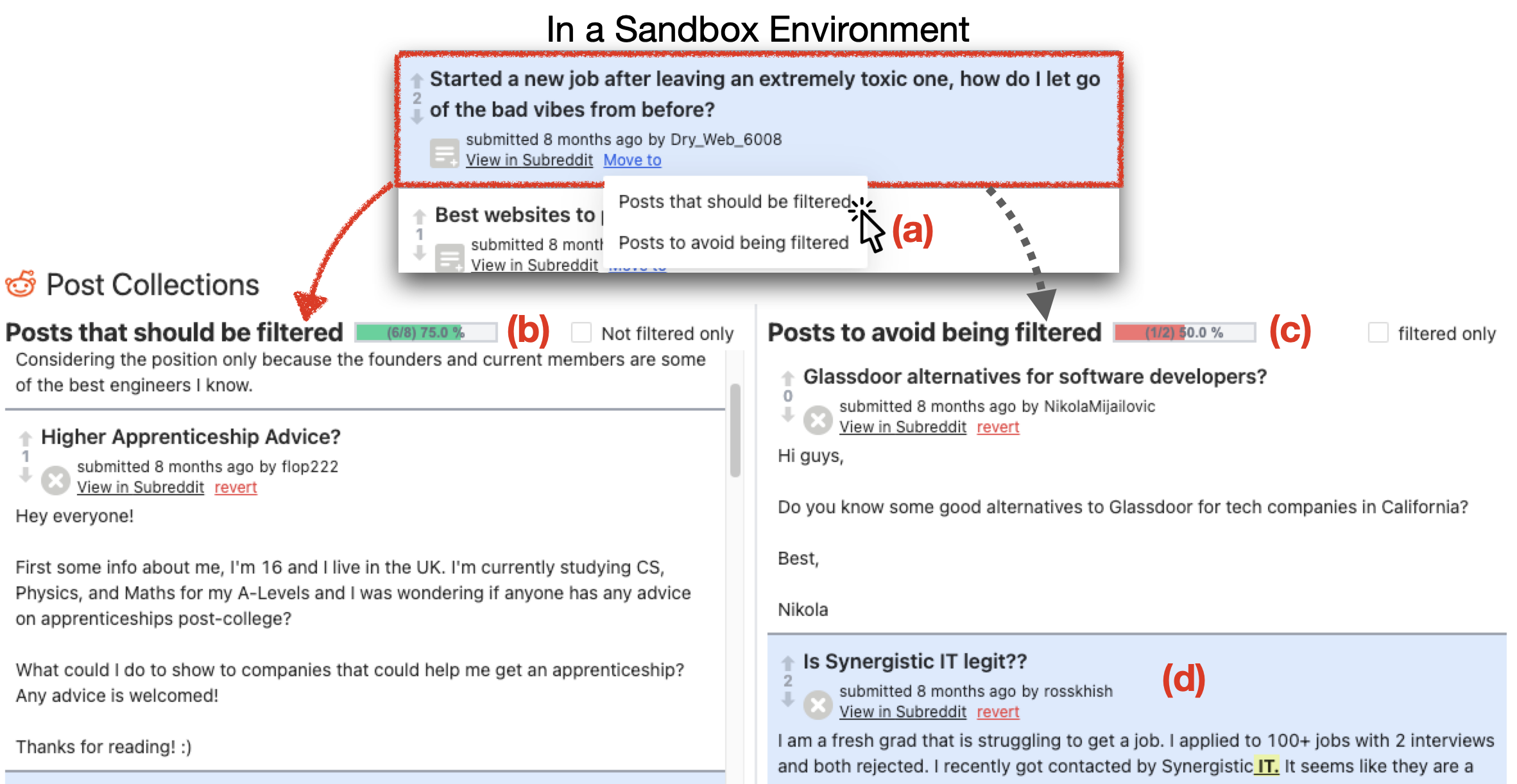}
    \caption{
        Show how to use FP/FN Collection. (a) The users can move posts from the Sandbox Environment to one of the Post Collections panels: ``Posts that should be filtered (red solid arrow)'' and ``Posts to avoid being filtered (gray dashed arrow)''. (b, c) The green and red bars show the ratio of the filtered ones. (d) The filtered posts by the automated rules are marked blue in the Post Collections panel.
    }
    \Description{This figure described how to move posts in a sandbox environment to FP/FN Collection, and explain UI components with alphabet labels}
    \label{fig:post_collections}
\end{figure}

The ``FP/FN Collection'' panel (\BC{3} Post Collections in Figure~\ref{fig:system}) enables moderators to collect posts that are useful for evaluating their rules, such as \textit{posts that should be filtered} (identified false negatives) or \textit{posts to avoid being filtered} (identified false positives)
Figure~\ref{fig:post_collections} shows an example of using the FP/FN Collection. A moderator can move the posts they want to filter with automated moderation to the ``Post that should be filtered'' panel ((a) in Figure~\ref{fig:post_collections}). If the community members are active at reporting the posts, moderators can put the reported posts right into the panel. Once enough posts are collected, the moderator can use this panel in two ways. First, they can browse through the posts to find patterns that could be useful to write an automated rule, e.g., find common keywords among the posts collected. Second, they can see a green bar to see the percentage of collected posts that are being filtered by the current automated rule ((b) in Figure~\ref{fig:post_collections}). If the number of posts being filtered is too low, they may want to update the automated rule to filter more posts. 

A similar practice could be applied to using the ``Post to avoid being filtered'' panel. A moderator can collect posts that should not be filtered in this panel to find common patterns among them. Then they can write an automated rule that would avoid filtering these posts. The moderator can monitor the red bar ((c) in Figure~\ref{fig:post_collections}) in this panel to see if the current rule is successfully avoiding filtering posts in this panel. For example, in this figure, since 50\% of the posts in this panel are being filtered, the moderator might want to edit their automated rule to reduce this number.

\begin{figure}[t]
    \centering
    \begin{minipage}{\textwidth}
    \centering
    \includegraphics[width=\textwidth]{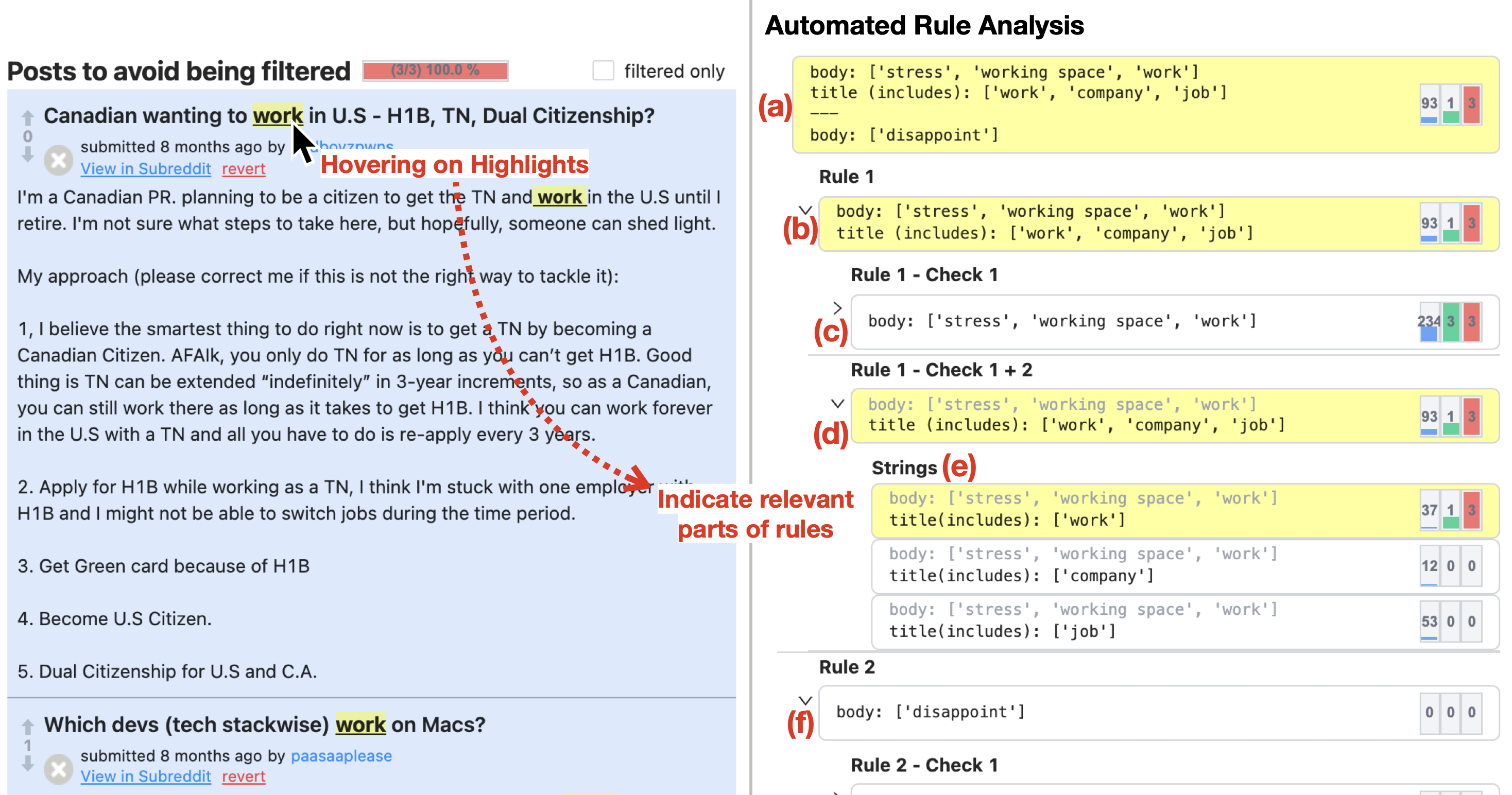}
    \caption{An example of the Automated Rule Analysis feature. Each labeled box with rounded corners on the right side represents a part of a configured rule. The three embedded vertical bar graphs on the right side of each rounded box show the number and ratio of filtered posts in three different types of posts: ``Posts on Subreddits'', ``Posts that should be filtered'', and ``Posts to avoid being filtered'', respectively.%
     (a) shows an AutoModerator configuration that consists of multiple rules: (b) and (f). Rule (b) detects intersection of posts detected by checks (c) and (d). Check (c) finds the posts that have any of `stress', `working space', and `work' in the body. Among the posts detected by check (c), check (d) detects the posts that includes any of `work', `company', and `job' on the title. (e) shows the impact of individual strings in the check (d). %
     The Highlight feature emphasizes specific part of posts affected by the configuration. As shown in the left side of the figure, when a user hovers the cursor on the word ``work'' in the post title, relevant items (the rounded boxes) are highlighted on Configuration Analysis panel. In this case, the first string in (e) got involved in the detection of ``work'' in the title. Thus, the system highlights the check (d), rule (b), and whole configuration (a) that includes the first string in (e)}
     \Description{This figure shows the user interface of Automated Rule Analysis features. It has alphabet labels on each item of automated rules, and show how mouse hovering works in the feature.}
     \label{fig:feature4}
     \end{minipage}\hfill
\end{figure}

\subsection {Feature 4: Automated Rule Analysis (DG4)}

\sys{} helps moderators analyze the impact of their complex automated rules through the features of ``Automated Rule Analysis'' (\BC{4} in Figure~\ref{fig:system}). First, ``Automated Rule Analysis'' panel shows rules in a hierarchical structure, allowing moderators to easily analyze them one by one. Similar to other automated rule generators, %
AutoModerator supports multiple rules, and each rule has one or more checks. A check is a line of code that represents a single condition for filtering the posts. It consists of an attribute of the posts, such as body or title, and a single list of strings. For example, \textit{body: [`red', `blue']} makes a rule to catch the posts with body that includes `red' or `blue'. Using the ``Configuration Analysis'' feature, moderators can assess the impact of each rule, check, and string individually. The three bar graphs in blue, green and red on the right panel of Figure~\ref{fig:feature4} indicate how each part of the rules affects posts in ``Posts on Subreddits'', ``Posts that should be filtered'' and ``Posts to avoid being filtered'', respectively. %
Looking at the three bar graphs in Figure~\ref{fig:feature4}(d), the rule 1 filter more posts in the ``Posts to avoid being filtered (red)'' than the posts in ``Posts that should be filtered (green)''. 
In this case, the moderator can expand Rule 1 for a deeper analysis by clicking on it and find that the keyword `work' in Rule 1 ((e)in Figure~\ref{fig:feature4}) is the one causing a lot of unwanted posts to be filtered. Then, they may choose to remove or update that keyword to reduce false positives.

``Automated Rule Analysis'' also presents quick highlights for all filtered posts showing which part of the post is being affected by the automated rules (e.g., the word in the post that triggers the AutoModerator) and which part of the rule is being triggered by the post (e.g., the keyword that was triggered in the filter). This feature helps moderators quickly and easily find the reason for false positives from automated rules. %
For example, if an automated rule is set to filter posts containing the word ``work'' in the body, then in every post that contains the word ``work'', the word ``work'' is highlighted in yellow (see Figure~\ref{fig:feature4}). Reversely, when a user hovers a cursor over one of the highlighted words in a post, the system highlights the triggered rule, check, and string so that the human moderator can understand which part of the rule is related to filtering the post. %

\section{User Study}

\begin{figure}[t]
    \centering
    \includegraphics[width=\textwidth]{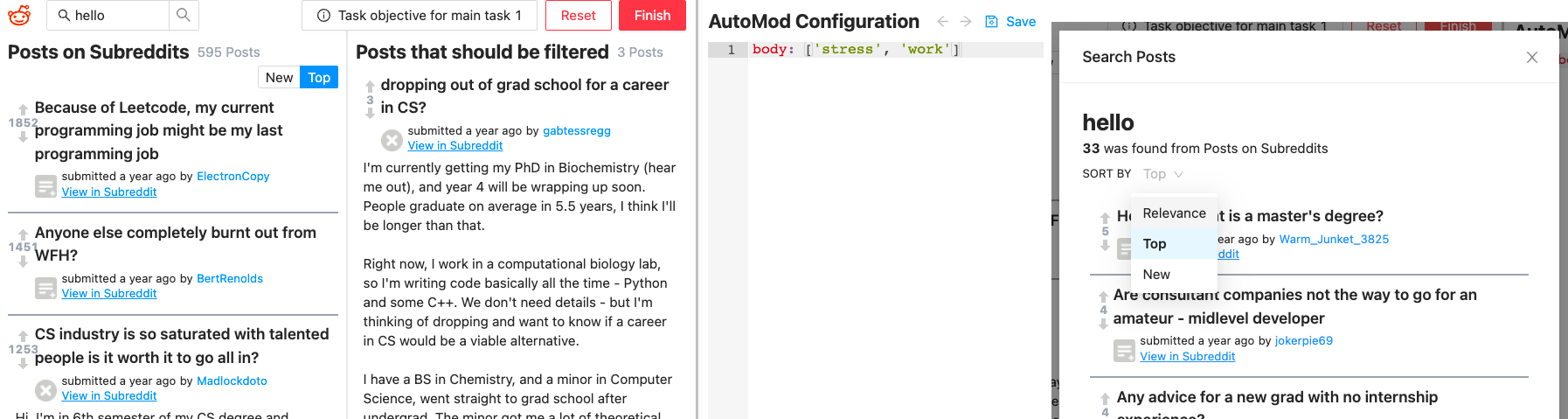}
    \caption{
        An overview of a basic system for our user study. The system provides similar features that moderators have during the moderation process in Reddit. The participants can see the community posts in ``Posts on Subreddits'', sort them by New \& Top, see the example target posts in ``Posts that should be filtered'', and search posts by words or phrases with the pop-up window in the right side.
    }
    \Description{This figure shows the overview of the basic system used in user study and pop-up window interface to search posts.}
    \label{fig:basic_system}
\end{figure}

To observe how \sys{} can change the AutoModerator configuration process, we conducted remote user studies with 10 active moderators of the online community through Zoom~\footnote{https://zoom.us/}. We first built a basic system (Figure~\ref{fig:basic_system}) that simulates a general process of creating Reddit AutoModerator rules. With the basic system, users can do the typical things they would do when moderating their subreddits: browse community posts, search posts by words or phrases, and sort posts by newest and highest votes. Then we built \sys{} by adding the features proposed in Section~\ref{sec:section4_new} to the basic system.

The posts to be used in the user study were crawled from a subreddit named r/cscareerquestions from May 1st 2021 to May 7th 2021, a subreddit where members post questions about computer science careers.
Our criteria to select a subreddit included (1) whether the community is active, (2) whether posts are mostly text-based as our scope focuses on keyword-based moderation, and (3) whether it is easy to make a plausible hypothetical community rule to use for the user studies. 

Through the user study, we aimed to evaluate whether \sys{} could help moderators identify the possible errors of AutoModerator and improve their automated rules. Furthermore, our goal was to analyze their process, perceived usefulness, and feedback to propose the direction for the future system that supports the automated moderation tool configuration process. 
\begin{itemize}

\item RQ1: Can ModSandbox support the configuration process of the automated moderation tool?
\begin{itemize}
    \item RQ1-1: Can ModSandbox support moderators to detect false positives and false negatives more easily?
    \item RQ1-2: Can ModSandbox help moderators update automated rules to reduce false positives and false negatives?
\end{itemize}
\item RQ2: How does ModSandbox support the configuration process of the automated moderation tool?
\begin{itemize}
    \item RQ2-1: How do participants use the features of ModSandbox for the configuration process?
    \item RQ2-2: How do participants perceive the usefulness of \sys{} in the configuration process?
\end{itemize}

\end{itemize}

\subsection{Participants}
We recruited 10 participants (Table~\ref{participants}), seven Reddit moderators (five males and two females) were from the United States. The other three were non-Reddit moderators (three females) and in charge of Korean online communities. These three were proficient in English, thus participated in the English-based user study as the U.S. participants did. We sent a recruitment advertisement to Reddit moderators through mod mail, which is a message system within the Reddit platform. We contacted moderators of subreddits randomly sampled from the same list for our interview recruitment and excluded the interview participants. The non-Reddit moderators in South Korea were recruited by word of mouth. We expected the non-Reddit moderators to represent voluntary moderators outside Reddit. Although they may not be familiar with Reddit, we confirm their moderation practices and challenges aligned with those on Reddit, while they might have unique moderation experiences. 

Additionally, we ensure that we have moderators both with and without experience using AutoModerator. Five of the participants (P1, P5, P7, P8, P10) had experience configuring AutoModerator while the others (P2, P3, P4, P6, P9), including Korean community moderators, had little or no AutoModerator experience. The recruitment method and the study design were approved by our institution's IRB policy.

\begin{table}[]
\resizebox{\textwidth}{!}{
\begin{tabular}{cccc|ccc|c}
 &  &  &  & \multicolumn{3}{c|}{\textbf{Prior experience}} & {\color[HTML]{333333} } \\ \cline{5-7}
\multirow{-2}{*}{\textbf{No.}} & 
\multirow{-2}{*}{\textbf{Age}} & \multirow{-2}{*}{\textbf{Gender}} & \multirow{-2}{*}{\textbf{Moderator periods}} & \textbf{Platform} & \textbf{AutoModerator} & \textbf{Programming} & \multirow{-2}{*}{{ \textbf{\begin{tabular}[c]{@{}c@{}}Condition for \\ User Study\end{tabular}}}} \\ \hline
P1 & 35-44 & M & over 5 years & Reddit & occasionally & basic concepts & Experienced \\ \hline
P2 & 18-24 & F & 6 months - 1 year & Non-Reddit & never & basic concepts & Novice \\ \hline
P3 & 25-34 & F & under 6 months & Non-Reddit & never & No knowledge & Novice \\ \hline
P4 & 18-24 & F & 6 months - 1 year & Non-Reddit & never & frequently & Novice \\ \hline
P5 & 25-34 & F & 1 - 2 years & Reddit & occasionally & basic concepts & Experienced \\ \hline
P6 & 25-34 & M & under 6 months & Reddit & done it once & No knowledge & Novice \\ \hline
P7 & 25-34 & M & 2 - 3 years & Reddit & occasionally & frequently & Experienced \\ \hline
P8 & 18-24 & M & 2 - 3 years & Reddit & occasionally & a few programs & Experienced \\ \hline
P9 & 45-54 & M & 6 months - 1 year & Reddit & done it once & a few programs & Novice \\ \hline
P10 & 18-24 & M & 1 - 2 years & Reddit & most of the time & a few programs & Experienced \\ \hline
\end{tabular}
}
\caption{Background Information of Study Participants. Experienced participants in \textbf{Condition for user study} are moderators who have experienced the configuration of AutoModerator occasionally or most of the time.
}
\Description{The table shows the background information (age, gender, moderator periods, prior experience, and user's condition) of user study participants}
\label{participants}
\end{table}

\subsection{Study Procedure}
Each study lasted about two hours, and each participant received a \$30 Amazon gift card per hour or 30,000 KRW per hour as compensation. Before the study, the participants filled out a consent form and answered their background information in Table~\ref{participants}

\subsubsection{Tutorial on How to configure AutoModerator (20-30 minutes)}

Before entering the main task, we explained what is expected in the main tasks. We also reviewed how to write AutoModerator rules that are directly related to the study. Next, we gave a walk-through tutorial on how to use the features in basic system and \sys{}, where the participants first used the basic system to get used to the task and wrote automated rules for the task, and then moved on to \sys{} to improve the rule using our proposed features. %

After the tutorial session, we asked participants to solve quizzes about the study to make sure that they understood the purpose of the study, how to configure AutoModerator, and how to use the systems. If they got incorrect answers, we helped them find the right answer and then checked if they understood correctly. This step ensured that everyone had rule authoring skills that were sufficient to perform the main tasks. We also provided participants with two reference documentations: the description of \sys{} features and the AutoModerator rule syntax, which they can freely access during the main tasks.

\subsubsection{Main Tasks (60-80 minutes)} \label{sec:tasks}
Each participant was given two different tasks where they write the AutoModerator rules for a given hypothetical moderation scenario.
Moderation according to the actual rules can expose users to mentally abusive posts including slurs and swear words. Thus, we created a novel moderation objective instead of using the subreddit's actual rules.

The two main scenarios that we showed to the participants were as follows:

\begin{itemize}
    \item Task A: Many people without CS relevant degrees post questions asking whether or how to get CS relevant jobs on r/cscareerquestions. Because r/cscareerquestions has a FAQ page that contains answers to those questions, moderators want to configure AutoMod to automatically leave a comment with a link to the FAQ page on posts asking whether and how to get CS-relevant jobs without the related degrees. \textbf{Objective:} Write AutoMod rules to detect posts asking whether or how to get CS-relevant jobs without CS-relevant degrees.
    \item Task B: The moderators of r/cscareerquestions want to leave a comment saying ``Your post includes keywords related to Covid-19. If you need any help with the current global pandemic situation related to medical, mental, or economical crisis, please contact xxx for further information.'' on posts relevant to ``covid''. \textbf{Objective:} Write AutoMod rules to detect the posts that the moderator should leave comments according to the above.
\end{itemize}

These two tasks represent two different scenarios of content moderation in online communities. The first task (CS-relevant degrees) represents a more community-specific moderation scenario, where the rule only applies to the specific community alone. This scenario also represents the cases where the targeted posts have semantically similar content, which makes it easier for natural language processing models to work.
The second task (COVID-19) represents a more general scenario in which unexpected external events affect the community.

For each task, we provided three target example posts as samples that meet each moderation objective. The participants were informed that those three example posts had already been manually filtered by other virtual peer moderators. Because moderation task is somewhat subjective, we expect that the given example posts would help participants have similar criteria on how they evaluate whether a post should be moderated or not. The authors selected this type of example posts from the posts that two external annotators regarded as targeted for the moderation objective. This process is further explained in Section~\ref{sec:GT_annotation}.  %

Each participant first used the basic system to draft automated rules and then moved on to \sys{} to improve the rules using the given features of the system. Task A and B were offered in a randomized order for each participant. %
To ensure that participants have reasonable rules to start with, we emphasized that the rules written in the basic system should be in the form of their best attempt before using \sys{}. 
To analyze how participants use \sys{} during the main task, their monitor screens were shared and then recorded with their consent.

\subsubsection{Post Surveys (10-20 minutes)}
After the main tasks, the participants took part in a survey about their experience. They answered a 7-point Likert scale and open-ended questions on how useful the features of \sys{} were in each main task, as well as the overall usefulness of \sys{}. We also asked them about their strategies using \sys{} and feedback on how the system could be improved.

\subsection{Measures}
To answer the research questions, we observe the following:

\emph{The distribution of target posts to be filtered within other posts under different sorting conditions. (RQ1-1)} This shows how much more effective ``FP/FN Recommendation'' (FP/FN) is than sorting by newest and highest votes (NEW and TOP). We expected our system to help users find false positives and false negatives more easily by showing more probable false positives and negatives on top. Therefore, we visualized the distribution of the target posts to compare our recommendation algorithm with the default ones. For comparison, we created a set of ground-truth (GT) target posts to be filtered for the two main tasks. The detailed procedure on obtaining these GT target posts is described in Section~\ref{sec:GT_annotation}.

\emph{The semantic similarities between the example posts and filtered posts. (RQ1-2)}
Since we provided example posts that represent each moderation goal, we can use the semantic similarities of the posts with the example posts to represent how each post is semantically close to false positives and false negatives. For example, if the filtered posts are semantically far from the example posts, they are likely to be false positives (or vice versa). Thus, we compare the similarities between example posts and filtered posts in the basic system and ModSandbox. If the system helps reduce false positives, their distribution will increase. We applied the algorithm that was used in ``FP/FN Recommendation'' feature to calculate semantic similarities.

\emph{The average complexity of the automated rules using \sys{}. (RQ1-2)} This shows how \sys{} can help participants build more sophisticated rules to reflect their moderation intention and avoid false positives and false negatives. We compared the number of rules, checks, and strings they wrote in the basic system and \sys{}. The number of rules can describe how many subgoals they considered for a given moderation scenario. The number of checks and strings can represent how accurately the rules catch the posts that the participants intended. 

\emph{System usage patterns of participants  and answers to a rule-making strategies (RQ2-1)}
Two authors reviewed screen recordings to observe how participants use a basic system and \sys{} and found patterns of using \sys{} features together to improve their automated rules. Also, we asked their own rule-making strategies while using \sys{} through the post surveys.

\emph{Perceived usefulness of each features (RQ2-2)}
We calculated the average usefulness score of \sys{} and its features for each task. We then analyze their answers to open-ended questions to understand why they gave those scores.

Finally, we directly asked for their feedback to improve \sys{} to set the direction for the future system.

\subsubsection{Creating a set of Ground-truth Target Posts to be Filtered}\label{sec:GT_annotation}
We hired two external annotators to label posts must be filtered for the given moderation scenarios. The posts to be filtered were labeled 1 and the posts not to be filtered were labeled 0. The inter-rater reliability measured with Cohen's Kappa was 0.45 for Task A and 0.67 for Task B. The scores were low even after having an asynchronous discussion session via email to reach agreement. This was because each annotator had different internal criteria for each scenario. For example, Annotator 2 considered that any post that mentions the usefulness of enrolling in a ``bootcamp'' should be filtered in Task A, while Annotator 1 did not agree with it. Although task B was more objective, the annotators still had disagreements between their labels. %
For example, Annotator 1 considered that any post that mentions ``lockdown'' should be filtered in Task B, while Annotator 2 did not agree with it. 
Thus, we did not directly compare user study participants configuration results with the ground truth dataset. The ground truth dataset was only used to assess the performance of the sorting algorithm.

\subsection{Results}

\subsubsection{RQ1-1: Can ModSandbox support moderators with detecting false positives and false negatives more easily?}

\begin{figure}[t]
    \centering
    \includegraphics[width=0.9\textwidth]{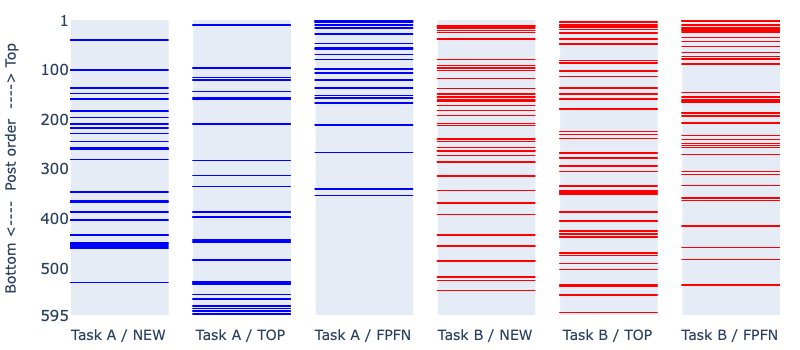}
    \caption{
        Locations of target posts to be filtered using different sorting methods in Task A and Task B. The target posts to be filtered labeled by two external annotators are marked in colors. \emph{FP/FN sorting} for Task A (the third column) concentrates the target posts to be filtered at the top of the list so that the users can more easily see them. 
    }
    \Description{}
    \label{fig:fpfn_sorting}
\end{figure} 

To visualize how ``FP/FN recommendation'' feature works on the main tasks of user study, we compared the order of posts between the \emph{NEW sorting}, \emph{TOP sorting}, and \emph{FP/FN sorting}. 
The results of these different sorting methods are shown in Figure~\ref{fig:fpfn_sorting}. We used the labeled dataset we created with two external annotators (Section~\ref{sec:GT_annotation}) to track the target posts to be filtered. Each post is marked as a line in blue and red. 
We found that the performance of FP/FN recommendation varies according to the moderation tasks. For Task A of filtering posts asking about getting CS-relevant jobs without CS-relevant degrees, many target posts to be filtered were located at the top when using \emph{FP/FN sorting}, and thus were first shown to the users (the third column in Figure~\ref{fig:fpfn_sorting}). This contrasts with \emph{new sorting} or \emph{top sorting} (the first and second column in Figure~\ref{fig:fpfn_sorting}). However, for Task B of filtering posts mentioning about COVID-19, \emph{FP/FN sorting} did not show noticeable differences from other sorting methods. %

\subsubsection{RQ1-2: Can ModSandbox help moderators update the automated rules to reduce the false positives and false negatives?}

\begin{figure}[t]
    \centering
    \includegraphics[width=0.9\textwidth]{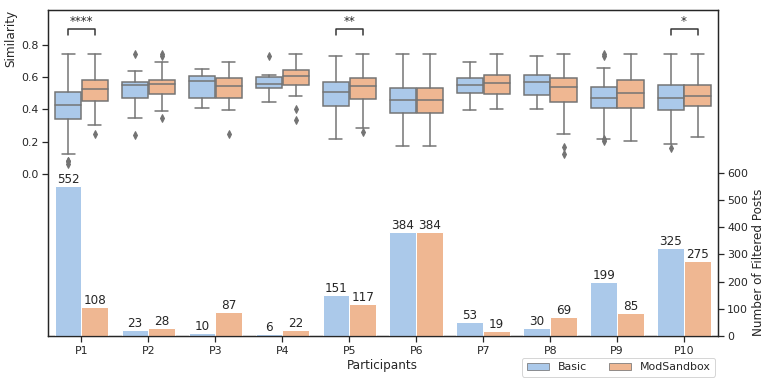}
    \caption{
    The bar graphs show the number of filtered posts and the whisker plots shows semantic similarity between filtered posts and target example posts for each participant on Task A.}
    \Description{The figure is a combined graph of bar graphs and whisker plots}
    \label{fig:filtered_similarity}
\end{figure}

We compared the characteristics of filtered posts after using the basic system and after using \sys{}. The results are shown in Figure~\ref{fig:filtered_similarity}, where the blue bars at the bottom indicate the number of posts filtered using the basic system, and the orange bars at the bottom indicate the number of posts filtered using \sys{}. The box whisker plots at the top shows the similarity of these posts compared to the three example posts which represents the targeted posts to be filtered. It is observed that for P1, P2, P4, P5, P7, P9, and P10, the similarity increased after using \sys{}. The difference of P1, P5, and P10 who are experienced AutoModerator users were statistically significant when tested with t-test. (p=0.000, p=0.004, p=0.033)  This implies that using \sys{}, the participants not only reduced the number of posts being filtered, but indeed the remaining filtered posts had high similarity with the targeted posts to be filtered.

\begin{table}[t]
  \centering
  \begin{tabular}{|c||c|c||c|c|}
  \hline
    & \multicolumn{2}{c||}{Task A} & \multicolumn{2}{c|}{Task B}\\
  \hline
  Unit & Before & After & Before & After\\
  \hline
  Number of rules & 1.4 (1.26) & \textbf{2 (1.86)} & 1.4 (0.70) & \textbf{1.6 (0.70)}\\
  Number of checks & 2.2 (2.44) & \textbf{3.7 (3.47)} & 1.8 (1.03) & \textbf{2.6 (1.96)}\\
  Number of strings & 10.6 (9.24) & \textbf{16.8 (14.02)} & 5.9 (3.73) & \textbf{8.1 (3.75)}\\
  \hline
  \end{tabular}
  \caption{Average number (and standard deviation) of rules, checks, and strings in the automated rules configured before using \sys{} and after using \sys{}}
  \Description{The table shows average number (and standard deviation) of rules, checks, and strings in the automated rules configured before using \sys{} and after using \sys{}}
  \label{tbl:rule_complexity}
\end{table}

We computed the average complexity of the first rules and the final rules by counting the number of rules, checks, and strings. We observed that participants start with relatively simple rule and make it to be more sophisticated as they use \sys{} (Table~\ref{tbl:rule_complexity}). Starting with the basic system, most participants made a single list of keywords and phrases rather than advanced combinations of units. Specifically, five moderators from Task A and nine moderators from Task B submitted a single-rule and single-check configuration. For example, P5 submitted a rule that detects any of 'change', 'degree', 'machine learning' and 'worth' in the posts for Task A. We also note one exceptional case where one moderator (P10) submitted a five-rule and nine-check configuration in Task A. As 
seen in Table~\ref{tbl:rule_complexity}, the participants tend to update their primitive rules to have more rule, check, and string after using \sys{}. This difference shows that \sys{} lets participants try more sophisticated rules to target the posts they want to filter accurately.

\subsubsection{RQ2-1: How do participants use the features of ModSandbox for the configuration process?}
\paragraph{System usage pattern in \sys{}}

We found that they used \sys{} to evaluate and update AutoModerater rules in a structured way; they created several routines of using \sys{}'s features. 

First, all participants preferred to activate the FP / FN Recommendation feature (Feature 3) rather than sorting by \emph{NEW} and \emph{TOP} throughout the study session to help find the posts that are likely to be false positives or false negatives.
Their process to update rules mostly began with finding false positives and false negatives using the ``FP/FN Recommendation'' feature (Feature 3). 
The most popular process was as follows. Seven participants (P1, P4, P5, P6, P8, P9, P10) first reviewed possible misses and false alarms that our feature recommended to find false positives and false negatives. Next, they moved the identified false positives and negatives into the ``FP/FN Collection'' panel. Then, they updated the AutoModerator configuration to resolve the collected posts on the FP/FN Collection panel. 

We observed two patterns in collecting problematic posts (usually actual false positives and false negatives) in FP/FN Collection (Feature 2). A group of participants (P1, P5, P6, P8, P9, P10) collected several problematic posts at once and updated their rules by referring to all of them at once. 
However, some of them (P6, P8, and P9) failed to use this pattern because they could not find a breakthrough rule to resolve the collected false positives and false negatives at once. %
Otherwise, Two participants (P4, P5) tried to collect posts one by one in the ``FP/FN Collection'' panel (Feature 2), followed by a rule update after each collection. This resulted in fine-tuning the rules to resolve each and every post being collected in the ``FP/FN Collection'' panel. 
An exceptional pattern was observed from P2 and P7, where they did not use the ``FP/FN Collection'' panel at all and directly reflected the false positives she found from the Possible False Alarms Panel. P2 reported that it was cumbersome to move posts to the FP/FN Collection because she was able to just quickly deal with the false positives she found without having to move them. All participants(P2, P4, P5, P7) using this one-by-one strategy presented a common strategy to update rules, which is writing a filter with a large number of keywords at first and then adding white-list keywords to exclude false positives. 
Six participants (P1, P5, P7, P8, P9, P10) used the Automated Rule Analysis Panel (Feature 4) as an extra supporting tool to understand how the rules are working. Four of them (P1, P5, P9, P10) refer to this feature to quickly find which part of the rules are filtering the false positives and false negatives in ``FP/FN Collection'' panel. The other two (P7 and P8) checked the number of posts that each rule and keyword affected. They first checked whether their rules were catching too many posts or not by looking at the ratio bar in the Sandbox Panel and used mouse hovering on highlights to remove the relevant part of the rules. 

\paragraph{Rule-making strategies in \sys{}}

Participants elaborated their rule-making strategies while using \sys{}. Four participants (P2, P4, P6, P8) first created a rule with a large list of keywords to catch targeted posts and update the rules to reduce the false positives. To be specific, P2, P4, and P8 added some white-list keyword filters to exclude the false positives. P8 described \textit{``First I skimmed through the posts and the task for keywords that might be able to match what I want. I then checked the false positives to find additional keywords that I could add to the rules to reduce the amount of false positives''.} P6 followed the same strategy, but he told that the configuration became so muddled that he was getting too many false positives and regretted that he should have thought a lot more different than he had at the beginning in trying to filter his words. P7 introduced his impressive strategy. He first tried to think of a simple algorithm that can catch or reject targeted posts in this head, e.g., find all posts including word X and word Y, but not any posts including word Z. Then he thought of keywords that fulfill that logic.

\subsubsection{RQ2-2: How do participants perceive the usefulness of \sys{} in the configuration process?}

\begin{table}[]
\begin{tabular}{|ll||l|l|l|l||l|}
\hline
Condition & Task & Sandbox & Collections & FP \& FN & Analysis & System \\ \hline
\multicolumn{1}{|l|}{\multirow{2}{*}{Experienced}} & Task A & 4.8(1.8) & \textbf{5.2(1.9)} & \textbf{6.0(0.7)} & 5.2(0.8) & \multirow{2}{*}{\textbf{6.2(0.8)}} \\
\multicolumn{1}{|l|}{} & Task B & 4.6(1.8) & \textbf{6.2(0.8)} & 5.4(1.1) & 5.2(1.1) &  \\ \hline
\multicolumn{1}{|l|}{\multirow{2}{*}{Novice}} & Task A & 5.0(1.6) & 4.6(1.6) & \textbf{5.8(0.8)} & 5.0(1.6) & \multirow{2}{*}{5.4(2.1)} \\
\multicolumn{1}{|l|}{} & Task B & \textbf{6.0(1.5)} & 5.0(0.9) & 5.4(0.4) & 5.0(2.0) &  \\ \hline
\multicolumn{1}{|l|}{} & Total & 5.1(1.6) & 5.2(1.6) & \textbf{5.7(0.9)} & 5.1(1.4) &  \\ \hline
\end{tabular}
\caption{Average usefulness scores (and standard deviation) of each feature in \sys{} }
\label{tbl:usefulness_score}
\Description{The table shows average usefulness scores (and standard deviation) of each feature in \sys{} }
\end{table}

After the user study, we asked all participants how useful each feature was and how they think they can be improved. We summarize the responses, highlighting the difference in main strength of each feature according to the moderation task and user's condition.

\paragraph{Feature 1. A Sandbox Environment: Valuable to see what posts are being filtered}
Six participants (P2, P3, P4, P6, P8, P9) responded that a sandbox environment was valuable because they could see what posts are being filtered in real-time. P1 said that he gave a high score for this function because it can show the results of AutoModerator applied to the community without affecting the community. However, three participants (P4, P7, P9) said that the Sandbox UI showed so much data compared to Feature 3: FP/FN Recommendation, that they did not like it. P7 said \textit{``There are a lot of posts shown at a time, which makes it less useful when compared to the features with fewer posts shown.''}

\paragraph{Feature 2. FP/FN Collection: More useful to participants who were familiar with configuring AutoModerator}
Six participants(P1, P2, P6, P7, P8, P10) mentioned that seeing the posts manually gathered in the Post Collections panel was useful for the user study task. Specifically, P6, P7, and P8 noted its usefulness in finding proper rules. P6 reported \textit{``It was great to actually see which posts are false negatives and false positives so that it was easier to look for keywords that are more relevant to the current topic.''} However, Three novice moderators (P3, P6, P9) pointed out they are difficult to use and gave lower usefulness scores to this feature (Table~\ref{tbl:usefulness_score}). P9 said \textit{``It was useful in that is showed me how the keywords were being used but it left me wondering how to apply this.''}

\paragraph{Feature 3. FP/FN Recommendation: The most useful feature for everyone, but only works well if posts have semantic similarities}
Five participants (P1, P4, P5, P8, P10) liked this feature because it allowed them to grasp \textit{probable} false positives and false negatives, and thus quickly find \textit{actual} false positive and negative posts. P10 mentioned \textit{``The possible misses, false alarms was very helpful in showing what things I missed with my filter. It definitely saved me tons of time of scrolling through matches to find bad ones.''} However, P2, P3, and P10 felt that possible misses are less accurate in Task B. P3 wrote \textit{``This feature is so convenient, but I think there were many articles in the Possible Misses that did not seem to be included in the task''.} Interestingly, P7 doubted the accuracy of the algorithm \textit{``I'm unsure how good the algorithm is and I'd be afraid that focusing on these will miss important posts''.} They evaluated this function as less useful in Task B, but as most useful function overall (Table~\ref{tbl:usefulness_score}). We note that this feature was indeed less accurate in Task B because each post mentioning COVID-19 had very different semantic and context compared to Task A. Targeted posts in Task A shared similar topics, but targeted posts in Task B had varying topics.

\paragraph{Feature 4. Automated Rule Analysis and Highlights: More useful with more complex rules}

``Automated Rule Analysis'' panel helped Four experienced moderators (P5, P7, P8, P10) when they analyze the code and determine which rules or words are good and bad for the task. P5 answered that it is helpful to see how each code impacts on the filtered results, thus making it easier for them to remove keywords that were yielding too many false results. Two experienced moderators (P1, P8) stated that they were able to understand how the rules work but they did not feel the need to use the panel much. Interestingly, P1 suggested a novel way to utilize what is seen in the Automated Rule Analysis panel. He pointed out that it is easily readable data that could be presented to other moderators as evidence to discuss the flaws and strengths of each rule.
Four novice moderators (P2, P3, P4, P9) preferred ``Highlights'' because it helps notice where the keywords in the rules are. Furthermore, P4 felt confident that she could identify why certain keywords were filtered or not. 

\paragraph{Feedback to improve \sys{}}
Three moderators (P1, P4, P7) provided feedback to improve \sys{} through the post-survey. P1 and P4 mentioned that the UI could be more simplified so that novice or casual moderators could also easily use it. 
P4 and P7 suggested analyzing word frequency in the ``FP/FN Collection'' panel so that the most frequent words can be used as recommended keywords when writing keyword-based automated rules.

\section{Discussion}

In this work, we investigate the challenges that online content moderators faced when configuring automated moderation tools and presented a novel approach to help them quickly find false positives and negatives and improve their automated rules.
In the following, we discuss the impact of intelligent NLP algorithms that help find false positives and false negatives, the potential impact of recommending concrete methods on how to update automated rules, supporting efficient collaboration between moderators, reducing emotional labor for online content moderators using \sys{}, and \sys{} being a practical solution for other platforms beyond Reddit.

\subsection{The Impact of Intelligent Algorithms on Finding False Positives and False Negatives}

In our user study, the accuracy of the algorithms in detecting possible false positives and negatives had a significant impact on the trust of participants and the perceived usefulness of the system. The experimental result in Figure~\ref{fig:fpfn_sorting} shows that our algorithm was not as effective as Task A in Task B. Due to this, three participants commented on the unreliability of the given functionality. P7 reported distrust of the algorithm: \textit{``I'm unsure how good the algorithm is and I'd be afraid that focusing on these will miss important posts.''} %
While the targeted posts to be filtered in Task A asked to filter posts with a similar context asking about getting CS-related jobs, Task B asked to filter posts that contain any keyword related to COVID-19, which may appear in various different contexts. These posts could have any topic spanning from talking about the impact of anti-vaccine protests in the job market to having to work remotely due to quarantine. In addition, the Universal Sentence Encoder may not be suitable for Task B because it was pre-trained with sources from Wikipedia, web news, web question-answer pages and discussion forums before the COVID-19 pandemic~\cite{cer2018universal}. 

The algorithm we adopted to predict false positives and negatives calculates semantic similarities of posts in the level of sentences, not keywords. 
Therefore, other algorithms could be tested to see the impact on tasks similar to Task B in our user study. 
For example, a word embedding model~\cite{mikolov2013efficient} or a language model pre-trained with recent social media content~\cite{loureiro2022timelms} may be more effective for similar tasks. 
For the current system, we only use a single algorithm, but it may be possible to improve the system by supporting alternative algorithms to recommend possible false positives and negatives, and let moderators compare the performance between them and apply what works best for them.
Another way to improve the feature to find possible false positives and false negatives is to expand the range of imported data. \sys{} extracts the possible misses and false alarms from only posts on a subreddit. The system can potentially use posts from multiple similar subreddits that share similar norms~\cite{chandrasekharan2018internet}. A more significant number of post-data can help moderators make a concrete and preventive configuration because they can provide various examples that reflect prospective behavior from similar communities~\cite{centola2010spread}.

\subsection{Further Recommending Concrete Ideas on How to Update the Automated Rules} %

Going further from just showing the possible false positive and false negative posts to the users, recommending concrete action items on how to update the automated rule may be helpful to the users. 
During the user study, three participants (P3, P6, P9) found it difficult to extract meaningful patterns to be written in a rule when using the ``FP/FN Collection'' panel. They lacked ideas to update the rules using these patterns because they were unable to identify the appropriate keywords. As a solution, we can leverage the ``FP/FN Collection'' panel to suggest concrete directions to improve the configuration. In the study, two participants (P4, P7) suggested showing frequently occurring keywords and inverse frequency analysis, which is a method to measure how much information each word provides. This approach may help find useful keywords to improve the rules based on the collected posts. Furthermore, \sys{} can potentially suggest a single regular expression that detects these useful keywords.

The patterns of rule update observed in the user study can guide the design of recommendations for future automated rules. Some participants added keywords they found in the possible false positive examples as white-list keywords. Furthermore, P10 started with a single check with a list of keywords and then added an additional normal check or reverse check, which is a condition that the post must not meet. These structured procedures can become a framework to help guide the writing of a better AutoModerator configuration. That is, we believe that guiding moderators to make informed updates to their AutoModerator configuration is a promising next step. The system can potentially recommend effective rules based on keyword extraction results and rule-update patterns. However, such data-driven recommendations may sometimes suggest rules that humans cannot interpret. To overcome this, the system may list promising options for changing current rules so that moderators can build more accurate and interpretable rules.

\subsection{Facilitating Distributed Governance for Online Communities}
In the user study, P1, who moderates a high-traffic subreddit, said \sys{} \textit{``would not only allow for refinement of rules, but presentation thereof''}. P1 meant that one could use \sys{} to demonstrate the expected results of AutoModerator configurations to peer moderators during discussions that are conducted before any moderation decision. P1 suggested that such use of \sys{} can help casual moderators become more involved in the configuration process. 
A previous study~\cite{jhaver2019human} showed that only a few moderators actively configure AutoModerator due to its difficulty in learning how to use it. Therefore, they suggested that an automated system could be designed to make it easier for moderators to understand how to use it. Tools that can visualize moderation rules and their results, such as \sys{}, can be a promising solution to support many non-tech-savvy moderators to participate in automated tools. In addition, \sys{} can support them in learning to use a regular expression in the configuration by testing it in the sandbox environment. We expect this line of work to reduce the barriers for novice moderators by providing a learning opportunity. 

Furthermore, \sys{} has the potential to serve a team of moderators and community users in their distributed decision-making scenarios. We can extend \sys{} to support multiple moderators in collaborative writing of rules while discussing the expected impact of automated rules on their community. We could even give these capabilities to community users, increasing moderation transparency and awareness. Recent studies(\cite{zhang2020policykit, schneider2021modular,jhaver2021designing}) have introduced software infrastructures and strategies to support distributed governance for online communities. For example, moderators can run a poll to make a decision about a change in an automated rule, showing statistics from \sys{}, not the rule itself. In this way, \sys{} contributes a special purpose software infrastructure and governance layer for algorithmic moderation to this research thread.

\subsection{Reducing Cognitive Labor in Setting Up Automated Moderation Tools}
The feature ``FP/FN Recommendation'' can help reduce the cognitive labor of moderators when configuring automated tools. For Task A in our study, we observed that this feature could help participants identify false positives and negatives earlier without skimming through all posts imported into the system (Figure~\ref{fig:fpfn_sorting}). P10 mentioned that \sys{} saved time in finding problematic posts compared to scrolling through a large number of community posts. 
Furthermore, moderators can avoid being exposed to toxic and harassing posts during moderation by using this feature. Although the typical moderation process requires emotional labor for moderators because they are exposed to these toxic posts while skimming through posts~\cite{kiene2016surviving, roberts2016commercial, roberts2019behind, dosono2019moderation}, using the ``FP/FN Recommendation'' feature creates a separate space for moderators to focus only on posts related to the current moderation task. 
Facebook recently built an AI-supported moderation system to reduce the amount of posts paid moderators should review by automatically excluding obviously harmful content and first sorting ambiguous content~\cite{vincent20facebook}. In a similar vein, \sys{} also benefits volunteer moderators by reducing the number of posts that they need to review.%

\section{Conclusion}
This paper proposes \sys{}, a virtual sandbox system for online content moderation, which supports human moderators in predicting and preventing false positives and false negatives of automated rules for their communities (e.g., filtering innocent posts or missing posts that should be filtered). \sys{} was built by investigating the four main challenges that moderators face during the configuration of their automated rules. The four main features driven from and corresponding to each challenge help moderators analyze their current automated rules and improve them by referring to the patterns found from the collected targeted posts to be filtered. Our user study with community moderators from various platforms demonstrates that \sys{} can help configure automated rules that reflect the detailed intentions of the moderators. Features like ``FP/FN Recommendation'' can reduce cognitive labor in setting up automated moderation rules because human moderators do not have to be exposed to toxic posts while analyzing their rules. Potential extended use cases of \sys{} include supporting collaboration between moderators by sharing the results of the system and reducing the barriers for novice moderators by providing a learning opportunity inside \sys{} with real community data.

\end{document}